\documentstyle[aps,floats,epsfig,axodraw]{revtex}

\input epsf

\def\beq{\begin{equation}}

\def\eeq{\end{equation}}

\def\bea{\begin{eqnarray}}

\def\eea{\end{eqnarray}}

\def\bq{\begin{quote}}

\def\eq{\end{quote}}

\def\gappeq{\mathrel{\rlap {\raise.5ex\hbox{$>$}}

{\lower.5ex\hbox{$\sim$}}}}

\def\lappeq{\mathrel{\rlap{\raise.5ex\hbox{$<$}}

{\lower.5ex\hbox{$\sim$}}}}

\def\Rp{R_p}

\def\Rpv{R_p \! \! \! \! \! \! /~~}

\def\mnu{[m_{\nu}]_{ij}}

\def\msusy{M_S}

\def\ltap{\raisebox{-.4ex}{\rlap{$\sim$}} \raisebox{.4ex}{$<$}}

\newcommand{\onetwo}{\Delta m^2_{12}}

\newcommand{\onethree}{\Delta m^2_{13}}

\newcommand{\twothree}{\Delta m^2_{23}}

\newcommand{\sun}{\Delta m^2_{\rm solar}}

\newcommand{\atm}{\Delta m^2_{\rm atm}}

\newcommand{\ssqsun}{\sin^2 2\theta_{12}}

\newcommand{\ssqatm}{\sin^2 2\theta_{23}}

\newcommand{\dmui}{\delta^i_{\mu}}

\newcommand{\dmuj}{\delta^j_{\mu}}

\newcommand{\dbi}{\delta^i_{B}}

\newcommand{\dbj}{\delta^j_{B}}

\newcommand{\dlijk}{\delta^{ijk}_{\lambda}}

\newcommand{\dlpipq}{\delta^{ipq}_{\lambda'}}

\def\bea{\begin{eqnarray}}   

\def\eea{\end{eqnarray}}

\begin{document}

\vspace*{-1in}

\renewcommand{\thefootnote}{\fnsymbol{footnote}}

\begin{flushright}

LPT Orsay/02-71\\

SINP/TNP/02-22\\

CI-UAN/02-08T\\

\texttt{hep-ph/0208009} 

\end{flushright}

\vskip 5pt

\begin{center}

{\Large {\bf A general analysis with trilinear and bilinear R-parity
violating couplings in the light of recent SNO data}}

\vskip 25pt 

{\bf Asmaa Abada $^{1,}$\footnote{E-mail
 address: abada@lyre.th.u-psud.fr}, Gautam Bhattacharyya
 $^{1,2,}$\footnote{E-mail address: gb@theory.saha.ernet.in}, Marta
 Losada $^{3,}$}\footnote{E-mail address: malosada@uan.edu.co}

\vskip 10pt 
$^1${\it Laboratoire de Physique Th\'eorique, Universit\'e
de Paris XI, B\^atiment 210, 91405 Orsay Cedex, France} \\ $^2${\it Saha
Institute of Nuclear Physics, 1/AF Bidhan Nagar, Kolkata 700064, India
}\\ 
$^3${\it Centro de Investigaciones, Universidad Antonio Nari\~{n}o,
Cll. 58A No. 37-94, Santa Fe de Bogot\'{a}, Colombia}

\vskip 20pt

{\bf Abstract}

\end{center}

\begin{quotation}

  {\noindent\small

\vskip 10pt

\noindent

We analyse an extension of the minimal supersymmetric standard model
including the dominant trilinear and bilinear R-parity violating
contributions. We take the trilinear terms from the superpotential and
the bilinear terms from the superpotential as well as the scalar
potential. We compute the neutrino masses induced by those couplings
and determine the allowed ranges of the R-parity violating parameters
that are consistent with the latest SNO results, atmospheric data and
the Chooz constraint.  We also estimate the effective mass for
neutrinoless double beta decay in such scenarios.  \\

\noindent 
PACS number(s):~12.60.Jv, 14.60.Pq, 11.30.Fs 
}

\end{quotation}

\vskip 20pt

\setcounter{footnote}{0}

\renewcommand{\thefootnote}{\arabic{footnote}}




\newpage

\setcounter{page}{1}
\pagestyle{plain}

\section{Introduction}
It is firmly established that non-zero neutrino masses can provide a
solution to the observed solar \cite{SNO,superk_sun} and atmospheric
neutrino deficits \cite{superk_atm}. This requires two large mixing
angles (very recent SNO data \cite{SNO} favour the large mixing angle
solution, LMA, for the solar anomaly) and two hierarchical neutrino
mass squared differences ($\sun \ll \atm$).  We assume that the three
mass eigenstates are given by $m_{i} ~(i =1,2,3)$ and we parametrise
the rotation matrix from neutrino flavour ($f$) to mass ($i$)
eigenstates as: $V_{fi} =
R_{23}(\theta_{23})R_{13}(\theta_{13})R_{12}(\theta_{12})$.  The MNS
matrix (analogous to the CKM mixing matrix for the quark sector)
neglecting CP phases is given by \cite{Maki:1962mu}, \beq V_{fi} =
\left[
\begin{array}{ccc} c_{12} c_{13} & c_{13} s_{12} &s_{13} \\ -c_{23}
s_{12} - c_{12} s_{13} s_{23} & c_{12} c_{23} - s_{12} s_{13} s_{23} &
c_{13} s_{23} \\ s_{23} s_{12} - c_{12} c_{23} s_{13} & -c_{12} s_{23} -
c_{23} s_{12} s_{13} & c_{13} c_{23}
\end{array} \right],
\label{mixing}
\eeq where $c_{ij} \equiv \cos \theta_{ij}$ and $s_{ij} \equiv \sin
\theta_{ij}$. For the solar anomaly, which we take to be a consequence
of $\nu_e$-$\nu_\mu$ oscillation, the relevant mass squared difference
is $\onetwo = \sun$, while for the atmospheric case, the oscillation
being between $\nu_\mu$ and $\nu_\tau$, the relevant mass squared
difference is $\onethree \approx \twothree = \atm$. After the
inclusion of the SNO data, the MSW-LMA oscillation is the most
favoured solution with $\sun = (2.5 - 19.0) \times 10^{-5}
~\mathrm{eV}^{2}$ and $\ssqsun = 0.61 - 0.95$ \cite{hs}. The SuperK
atmospheric neutrino data suggest $\atm = (2 - 5) \times
10^{-3}~\mathrm{eV}^{2}$ with $\ssqatm = 0.88 - 1.0$
\cite{superk_atm}. The Chooz \cite{chooz} and Palo Verde \cite{palo}
long baseline reactor experiments bound $\sin^2
\theta_{13}~\ltap~0.04$. Concerning the absolute masses, the recent
claim for the evidence of neutrinoless double beta decay
($0\nu\beta\beta$) by the Moscow-Heidelberg Collaboration \cite{db},
although not yet firmly established, constrains the ($ee$)-element of
the neutrino mass matrix in the flavour basis to lie between 0.1 and
0.5 eV. Tritium $\beta$-decay requires $m_{\nu_e}~\ltap~2.2$ eV
\cite{tritium}. Cosmological analysis from the recent 2dF Galaxy
Redshift Survey constrains $\sum_i m_i~\ltap~2.2$ eV \cite{galaxy}.

Neutrino masses can be generated in the R-parity violating ($\Rpv$)
 Supersymmetric Standard Model, where $\Rp$ is defined as $(-1)^{3B+L
 +2S}$ \cite{rpar1,rpar2}. Here $B,L$ and $S$ are the baryon number,
 lepton number and spin of a particle, respectively.  Strict
 phenomenological bounds on $B$ and/or $L$ violation exist in the
 literature \cite{reviews}.  In this paper we assume that $B$
 violating couplings are absent and generate neutrino Majorana masses
 via two units of $L$ violation. For this purpose, we allow both
 bilinear ($\mu_i$) and trilinear ($\lambda, \lambda'$) interactions
 in the superpotential as well as the bilinear soft terms
 ($B_{i}$)~\footnote{We follow closely the conventions given in
 \cite{DL1,DL2,AMS} and we refer the reader to these papers for more
 details. The symbols used in Eqs.~(\ref{S}) and (\ref{soft}) have
 their usual significance.}, \beq W= \mu^J {H}_u L_J +
 \frac{1}{2}\lambda^{JK \ell} L_J L_K E^c_{\ell} + \lambda^{'Jpq}
 L_JQ_p D^c_q + h_u^{pq} {H}_u Q_p U^c_q \ ,\label{S} \eeq where the
 vector $L_J = (H_d, L_i)$ with $J:4..1 $, and the soft
 supersymmetry-breaking potential is
\begin{eqnarray}
V_{soft}& = & \frac{\tilde{m}_u^2}{2} H_u^{\dagger} H_u + \frac{1}{2}
 L^{J \dagger} [\tilde{m}^2_L]_{JK} L^K  + B^J H_u L_J   \nonumber \\ 
& & + A^{ups} H_u Q_p U^c_s + 
     A^{Jps} L_J Q_p D^c_s +
    \frac{1}{2}A^{JKl} L_J L_K E^c_l + {\rm ~h.c.}  \label{soft}
\end{eqnarray}

 As has been extensively discussed in refs \cite{GH,DL1,DL2,AMS} (and
 references therein), field redefinitions of the $H_{d}, L_{i}$ fields
 correspond to basis changes in $L_{J}$ space and consequently the
 Lagrangian parameters will be altered.  We use the basis-independent
 parameters constructed in \cite{DL1,DL2} and write the neutrino mass
 matrix in terms of $\dmui, \dbi, \dlijk, \dlpipq$, which in the basis in
 which the sneutrino vevs are zero correspond to $\mu^i/|\mu|$,
 $B^i/|B|$, $\lambda_{ijk}$, $\lambda'_{ipq}$, respectively.

Calculations of neutrino masses in the context of $\Rpv$
theories  have focused on tree-level contributions from the
bilinear $\mu_i$-parameter  or loop contributions from
the trilinear $\lambda$ and $\lambda'$ couplings \cite{everyone}. 
Recently a detailed phenomenological analysis has been done for a model
including both $\lambda$, $\lambda'$ and the superpotential bilinear
parameter $\mu_i$ \cite{AM1,AM2}. In the basis-independent approach, a
phenomenological analysis has been done including only the purely
bilinear contributions, $\delta_{\mu}^i$ from the superpotential and
$\delta_{B}^i$ from the soft Lagrangian \cite{AMS}. The possibility that
loops involving $\delta_{\mu}$ were important was also discussed in
\cite{AMS}.

Our purpose in this paper is to perform for the first time an analysis
in a model which includes both the superpotential and soft bilinear
parameters $\dmui$ and $\dbi$ along with the superpotential trilinear
couplings $\dlijk$, in the mass insertion
approximation. A similar analysis can be performed including the
trilinears $\dlpipq$.  In our analysis we take all $\Rpv$ parameters
to be real. Thus we update the analysis in \cite{AMS} by including the
trilinear loop contributions alongside the bilinear tree and loop
terms.  We interface the neutrino mass matrix constructed out of the
$\Rpv$ parameters with the constraints on the solar and atmospheric
mass squared splittings, the corresponding mixing angles, and the Chooz
constraint on $\theta_{13}$. We then examine the nature of the mass
spectrum and we check the consistency with cosmological data. We
also obtain the effective mass for neutrinoless double beta decay in
our scenario. Furthermore, we observe the impact of adding the
trilinear contributions to the numerical results obtained in
\cite{AMS}.

\section{Parametrization of the mass matrix}
The $\Rpv$ model we are considering generates a single neutrino mass
at tree level proportional to $\dmui \dmuj$.  There are additional
loop corrections to the mass matrix when non-vanishing $\dbi$,
$\dlijk$ (and $\dlpipq$) are included, leading to more than one
non-zero mass eigenvalue. This enables us to fit the data on mass
squared splittings and mixing angles.  The relevant types of loops we
consider in the mass insertion approximation are:
\begin{itemize}
\item the well-known loops involving the trilinear $\Rpv$ couplings
$\lambda$ or $\lambda'$ at the neutrino vertices I and II in Fig.~1
(with lepton/slepton or quark/squark as propagators), which give
contributions proportional to $\delta_{\lambda} \delta_{\lambda}$ (or
$\delta_{\lambda'} \delta_{\lambda'}$);

\item the Grossman-Haber diagrams \cite{GH}, in which there are gauge
couplings at the neutrino vertices while there are two types of $\Rpv$
interactions contributing to the $\Delta L = 2$ Majorana mass in the
diagram of Fig.~\ref{fGH}. The
first kind have $\Rpv$ couplings located at positions III $+$ IV
(slepton-Higgs mixing) with contributions proportional to $\delta_B
\delta_B$. In the second type, the $\Rpv$ interactions are located at
positions V $+$ IV (neutrino-neutralino and slepton-Higgs mixing)
with contributions proportional to $\delta_{\mu} \delta_B$;

\item the diagram of Fig.~\ref{fmulambda}, where two units of $L$
violation come from positions V (neutrino-neutralino mixing) and II
($\lambda$ or $\lambda'$ vertex). The contribution to the neutrino
mass is proportional to $\delta_{\mu} \delta_{\lambda}$ (or
$\delta_{\mu} \delta_{\lambda'}$).
\end{itemize}

The last two types of loops have frequently been overlooked in analytic
estimates of neutrino masses.  Exact formulae for these diagrams can be
found in \cite{DL2}. Our  formulae are robust (see later) and good enough
for an order-of-magnitude estimate.

\begin{center}
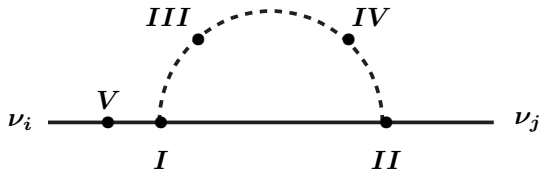
\begin{figure}[htb]
\unitlength1mm
\SetScale{2.8}
\begin{boldmath}

\begin{center}
\begin{picture}(60,50)(0,-10)
\Line(0,0)(15,0)
\Line(45,0)(15,0)
\Line(60,0)(45,0)
\DashCArc(30,0)(15,0,180){1}

\Text(8,0)[c]{$\bullet$}
\Text(8,3)[c]{$V$}
\Text(15,0)[c]{$\bullet$}
\Text(15,-5)[c]{$I$}
\Text(45,0)[c]{$\bullet$}
\Text(45,-5)[c]{$II$}
\Text(20,11)[c]{$\bullet$}
\Text(16,14)[c]{$III$}
\Text(40,11)[c]{$\bullet$}
\Text(43,14)[c]{$IV$}
\Text(-2,0)[r]{$\nu_i$}
\Text(62,0)[l]{$\nu_j$}
\end{picture}
\end{center}

\end{boldmath}
\caption{ The conventional loops ($\Rpv$ at I $+$ II) and the
Grossman-Haber loops ($\Rpv$ at III $+$ IV or V $+$ IV) contributing
to the neutrino mass.  }
\label{fGH}
\end{figure}

\begin{figure}[htb]
\unitlength1mm
\SetScale{2.8}

\begin{boldmath}
\begin{center}

\begin{picture}(60,60)(0,-20)
\Text(8,0)[c]{$\bullet$}
\Text(8,3)[c]{$V$}
\Text(45,0)[c]{$\bullet$}
\Text(45,-5)[c]{$II$}


\Line(0,0)(15,0)
\Line(45,0)(15,0)
\Line(60,0)(45,0)
\DashCArc(30,0)(15,0,180){1}
\Text(-2,0)[r]{$\nu_i$}
\Text(62,0)[l]{$\nu_j$}
\Text(33,-20)[c]{$v_I$}
\DashLine(33,0)(33,-15){1}
\end{picture}
\end{center}
\end{boldmath}
\caption{Charged loops with one gauge and a trilinear Yukawa
 coupling. The $\Rpv$ interactions are at V and II.}
\label{fmulambda}
\end{figure}
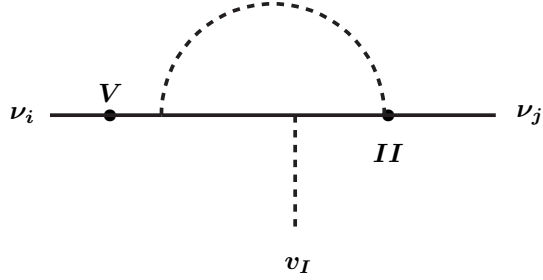

\end{center}

Setting all unknown sparticle masses equal to $\msusy$ = 100 GeV,
and neglecting the mixing angles among the sparticles, we have a
neutrino mass matrix of the form\newpage

\bea \mnu &=& \msusy \left[\dmui \dmuj +\frac{\kappa_1}{\cos\beta}
\left(\dmui \dbj + \dmuj \dbi\right) + \frac{\kappa_1}{\cos^2\beta}
\dbi \dbj \right] \nonumber \\ &+& \kappa_2 \left[\sum_{k,n} m_{e_{n}}
m_{e_{k}} \delta_{\lambda}^{ink}\delta_{\lambda}^{jkn} + 3 \sum_{k,n}
m_{d_{n}} m_{d_{k}}
\delta_{\lambda'}^{ink}\delta_{\lambda'}^{jkn}\right] \\ &+& \kappa_3
\left[\sum_{k} m_{e_{k}} (\dmui \delta_{\lambda}^{jkk} + \dmuj
\delta_{\lambda}^{ikk}) + 3\sum_{k} m_{d_{k}} (\dmui
\delta_{\lambda'}^{jkk} + \dmuj \delta_{\lambda'}^{ikk}) \right]
\nonumber, \eea where

\bea \kappa_1 = \frac{g^2}{64\pi^2},
~ \kappa_2 = \frac{1}{8 \pi^2 \msusy},~ \kappa_3 = \frac{g}{16\pi^2\sqrt
2}.  \eea We have included for
completeness the contributions arising from the $\delta_{\lambda'}$
terms which we put to zero in our numerical analysis.  The simplest
case to consider with a common $\delta_{\mu}^{i} \equiv
\delta_{\mu},\delta_{B}^{i} \equiv \delta_B $ and $
\delta_{\lambda}^{ink} \equiv \delta_{\lambda}$ does not work as it
gives only two non-zero masses and an eigenvector of the form
$(1/\sqrt 2,
-1/\sqrt 2 , 0)$ which cannot accomodate two large mixing angles for
$\theta_{12}$ and $\theta_{23}$.

In our numerical analysis we take : $\delta_{\mu}^{i}$,
$\delta_{B}^{i} $, $\delta_{\lambda}^{ink} \equiv \delta_{\lambda}$,
$\delta_{\lambda'}^{ink}=0$, for $i=$ 1, 2, 3, i.e., seven independent
parameters.  Thus, the neutrino mass matrix elements will be given by,
\bea m_{11} &=& \msusy\left[(\delta_{\mu}^{1})^{2} +
\frac{\kappa_1}{\cos^2\beta} (\delta_{B}^{1})^{2} +
2\frac{\kappa_1}{\cos\beta}\delta_{\mu}^1 \delta_{B}^1\right]+
2\kappa_3 m_{\tau} \delta_{\mu}^1\delta_{\lambda} +\kappa_{2}
m_{\tau}^{2} \delta_{\lambda}^{2}, \nonumber \\ m_{12} &=&
\msusy\left[\delta_{\mu}^{1}\delta_{\mu}^{2} +
\frac{\kappa_1}{\cos^2\beta} \delta_{B}^{1} \delta_{B}^{2} +
\frac{\kappa_1}{\cos\beta}(\delta_{\mu}^1 \delta_{B}^2 +
\delta_{\mu}^2 \delta_{B}^1)\right]+ \kappa_3 m_{\tau}
(\delta_{\mu}^1\delta_{\lambda} +\delta_{\mu}^2\delta_{\lambda})
+\kappa_{2} m_{\tau}^{2} \delta_{\lambda}^{2}, \nonumber \\ m_{22} &=&
\msusy\left[(\delta_{\mu}^{2})^{2} + \frac{\kappa_1}{\cos^2\beta}(
\delta_{B}^{2})^{2} + 2\frac{\kappa_1}{\cos\beta}\delta_{\mu}^2
\delta_{B}^2\right] + 2\kappa_3 m_{\tau}
\delta_{\mu}^2\delta_{\lambda} +\kappa_{2} m_{\tau}^{2}
\delta_{\lambda}^{2},  \\ 
m_{13} &=&
\msusy\left[\delta_{\mu}^{1}\delta_{\mu}^{3} +
\frac{\kappa_1}{\cos^2\beta} \delta_{B}^{1} \delta_{B}^{3} +
\frac{\kappa_1}{\cos\beta}(\delta_{\mu}^1 \delta_{B}^3 +
\delta_{\mu}^3 \delta_{B}^1)\right] + \kappa_3 m_{\tau}
\delta_{\mu}^3\delta_{\lambda}, \nonumber \\
m_{23} &=&
\msusy\left[\delta_{\mu}^{2}\delta_{\mu}^{3} +
\frac{\kappa_1}{\cos^2\beta} \delta_{B}^{2} \delta_{B}^{3} +
\frac{\kappa_1}{\cos\beta}(\delta_{\mu}^3 \delta_{B}^2 +
\delta_{\mu}^2 \delta_{B}^3)\right] + \kappa_3 m_{\tau}
\delta_{\mu}^3\delta_{\lambda}, \nonumber \\ 
m_{33} &=&
\msusy\left[(\delta_{\mu}^3)^{2} + \frac{\kappa_1}{\cos^2\beta}
(\delta_{B}^3)^{2} + 2\frac{\kappa_1}{\cos\beta}\delta_{\mu}^3
\delta_{B}^3\right], \nonumber\eea 
where we have employed the hierarchy of the
charged fermion masses to keep only the dominant terms.

\section{Results and Conclusions}
We have performed a general scan of parameter space made up by the
seven parameters that appear in the mass matrix allowing tree-level
contributions to either dominate over the loop corrections, to be on
the same order as these, or to be much smaller than the loop terms.
The fitted values of the couplings, obtained by using the atmospheric
and Chooz data together with the preferred solar MSW-LMA solution from
SNO, are presented in table I. Here we have taken $\cos\beta=1$
primarily to ensure a comparison with results in \cite{AMS}, which too
employed the same value of $\cos\beta$, on the same footing.  We
stress that although this choice of $\cos\beta$ leads to an
unacceptably low $\tan\beta$, we still use this value for illustration
and effective comparison with previous results, noting at the same
time that an order-of-magnitude estimate of the couplings is not
sensitive to this choice.  We also show in table I the allowed ranges
of $\sum_i m_i$, and $m_{\rm eff} \equiv \sum_i V^2_{ei} m_i = \sum_i
|V_{ei}|^2 m_i$ (since we have assumed the $V$ matrix to be real).
This last quantity is the effective mass relevant for neutrinoless
double beta decay (the neutrino masses induced by $\Rpv$ interactions
are Majorana type).

\begin{table}[htb]

\begin{tabular}{|l|l|l|}
Couplings & Min & Max\\\hline
$\delta_\lambda $ & $-2.0\times 10^{-4}$  & $2.0\times 10^{-4}$ \\\hline
$\delta_\mu^1$& $-6.8\times 10^{-7}$& $6.8\times 10^{-7}$ \\\hline
$\delta_\mu^2$& $-8.4\times 10^{-7}$& $8.4\times 10^{-7}$\\\hline
$\delta_\mu^3$& $-8.4\times 10^{-7}$& $8.4\times 10^{-7}$ \\\hline
$\delta_B^1$& $-2.7\times 10^{-5}$& $2.7\times 10^{-5}$ \\\hline
$\delta_B^2$& $-3.0\times 10^{-5}$& $3.0\times 10^{-5}$\\\hline
$\delta_B^3$& $-3.0\times 10^{-5}$& $3.0\times 10^{-5}$ \\\hline\hline
$m_{\rm eff} ({\rm eV})$ & $1.9\times 10^{-3}$& $6.7\times 10^{-2}$ \\
\hline
$\sum m_i ({\rm eV}) $ & $4.9\times 10^{-2}$& $0.2$ \\
\end{tabular}
\protect\label{seven}
\caption{Fitted range of the couplings satisfying MSW-LMA, SuperK and
 Chooz simultaneously. The fit is performed taking $\cos\beta=1$.}
\end{table}

\begin{figure}[htbp]

\hspace{3cm}

\hspace{2cm}\epsfxsize=14cm\epsfbox{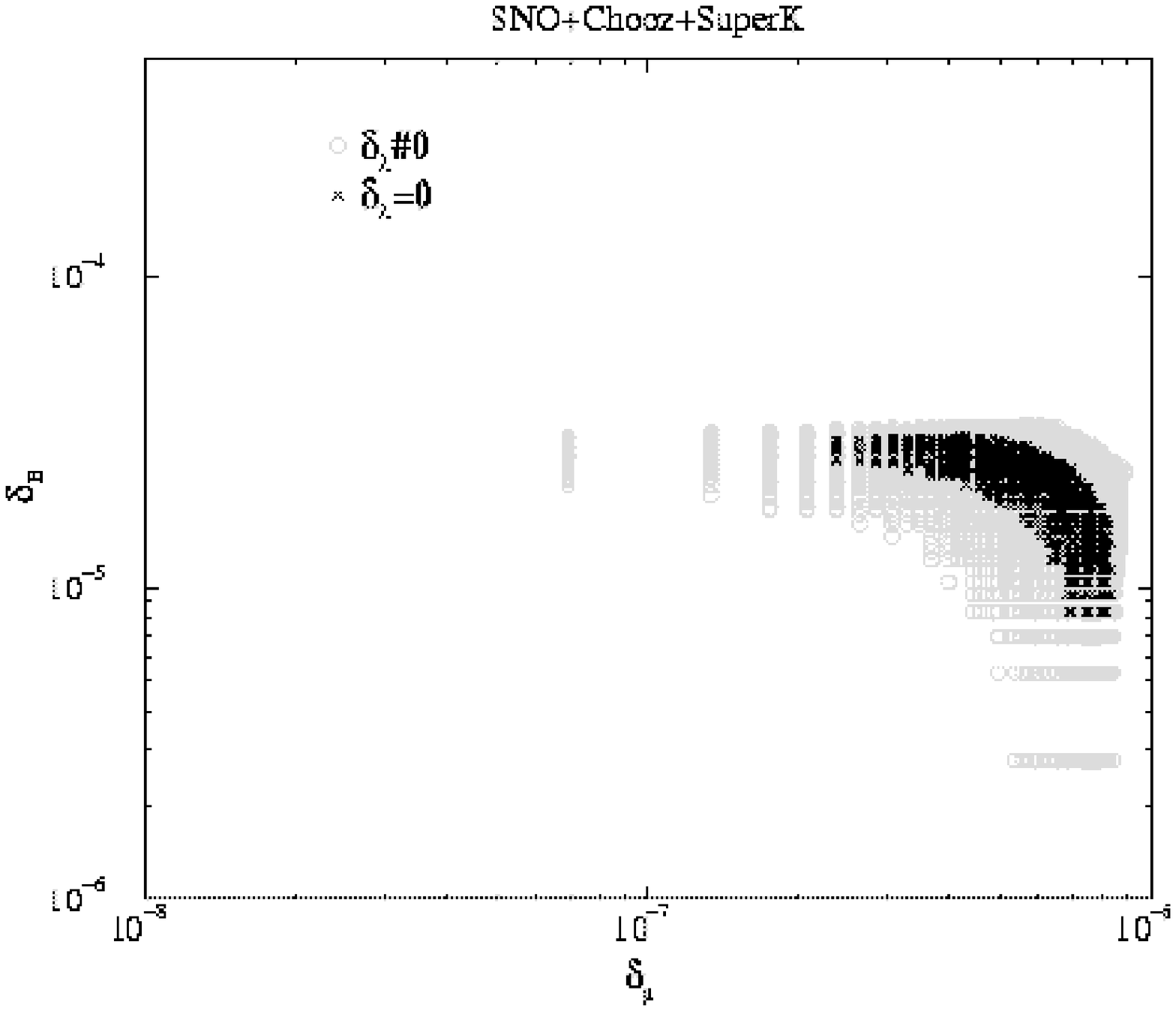}

\caption{ We present the available region in the $\delta_B \equiv
|\delta_B|$ versus $\delta_{\mu} \equiv |\delta_{\mu}|$ plane for a
combined fit using Chooz, SuperK and the SNO results for the MSW-LMA
solar oscillation solution, with $\cos\beta=1$. The circles are
solutions for $\delta_{\lambda} \neq 0$, the crosses are for
$\delta_{\lambda}=0$. This figure should be compared with Fig.~5 of
ref.~[17].}

\protect\label{dmuvsdB}

\end{figure}

In Fig.~\ref{dmuvsdB} we present the allowed region in the $|\delta_B|
= \sqrt{\sum_i (\dbi)^2}$ versus $ |\delta_{\mu}| = \sqrt{\sum_i
(\dmui)^2}$ plane for the combined fit taking $\cos\beta =1$. We
present our results for both $\delta_{\lambda} \neq 0$ and
$\delta_{\lambda}=0$.  It can be clearly seen that the allowed region
increases when we admit non-zero values of $\delta_{\lambda}$. This is
mainly due to the presence of the $\delta_\mu \delta_{\lambda}$ terms
in the mass matrix (originating from Fig.~\ref{fmulambda}) which can
take either sign and thus can accomodate a larger region of parameter
space.

The resulting fit strongly prefers a hierarchical mass pattern in our
scenario, although our analysis cannot make a distinction between the
inverted and the normal hierarchy.  The inclusion of a non-zero
$\delta_{\lambda}$ to the set of non-vanishing $\dmui$ and $\dbi$
allows us to have all three non-vanishing neutrino masses while if
only the latter two vectors are non-zero there are only two non-zero
eigenvalues \cite{AMS}.  The maximum value of $m_{\rm eff}$ we predict
can be tested in the next generation of neutrinoless double beta decay
experiments.  We have also checked that the cosmological bound
restricting the sum of eigenvalues to be at most $\sim 2$ eV is
fulfilled.

To conclude, we outline the novel features of our analysis: (i) the
trilinear $\Rpv$ interactions along with {\em both} types (originating
from superpotential and soft terms) of bilinear $\Rpv$ terms have been
employed for the first time to analyse the neutrino mass matrix; (ii)
the seven parameter neutrino mass matrix has been interfaced with the
latest SNO results favouring MSW-LMA solution along with the SuperK
atmospheric and Chooz data; (iii) by including a non-vanishing
$\delta_{\lambda}$ we have updated the analysis in \cite{AMS} and the
numerical impact of adding this term has been highlighted in
Fig.~~\ref{dmuvsdB}; (iv) the mass spectrum in this kind of scenario
implies a hierarchical pattern, although a distinction between the
normal and inverted pictures cannot be made; (v) our prediction of
$m_{\rm eff}$ can be tested in the next generation of neutrinoless
double beta decay experiments (GENIUS, Majorana, MOON, EXO) \cite{EV}.
We finally make a remark that allowing a non-vanishing
$\delta_{\lambda'}$ will not qualitatively change the pattern of our
fit.

\section*{Acknowledgements}
G.B. acknowledges hospitality at LPT, Univ. de Paris, Orsay, while
this work has been done.  The work of M.L. was partially suported by
Colciencias-BID, under contracts nos. 120-2000 and 348-2000.

\end{document}